\documentclass[preprint,12pt]{elsarticle}

\begin{document}

\begin{frontmatter}

\title{A Model of Opinion Dynamics with Bounded Confidence
and Noise}

\author{P. Nyczka}

\address{Institute of Theoretical Physics, University of Wroc{\l}aw, pl. Maxa
Borna 9, 50-204 Wroc{\l}aw, Poland }


\begin{abstract}
This paper introduces a new model of continuous opinion dynamics with random noise. The model belongs to the broad class of so called bounded confidence models. It differs from other popular bounded confidence models by the update rule, since
it is intended to describe how the single person can influence at the same time a group of several listeners. Moreover, opinion noise is introduced to the model. Due to this noise, in some specific cases, spontaneous transitions between two states with a different number of large opinion clusters occur. Detailed analysis of these transitions is provided, with MC simulations and ME numerical integration analysis.
\end{abstract}

\end{frontmatter}

\section{Introduction}
\label{sec:1}
Models of opinion dynamics are very popular in modern sociophysics (see recent reviews \cite{CFL07,G08,SW05}). An interesting subset of them are the models with Bounded Confidence (BC models) \cite{L07,DNAW00,HK02,WDA05}. In these models, the opinion exchange takes place only when the difference between two opinions is below the confidence bound (also called as threshold or tolerance). This is a reasonable consideration because if the minds of two people are very different, it is difficult for them to convince each other of something. Sometimes it is even hard for them to talk to each other. As a result of opinion exchange, one agent can change the opinion of another agent, or they can convince each other. There are many different types of opinion exchanges, including exchanges between more than two agents. BC models are commonly used to simulate the evolution of opinion distribution in a set of agents. Depending on tolerance, the simulation results can give consensus (one big cluster in an opinion space) for big Tolerance, polarization (two big clusters) for smaller Tolerance, or clusterization  (three or more clusters) for even smaller Tolerance.

Although several extensions based on continuous opinions and bounded confidence have been proposed and analyzed \cite{L07}, little attention has been paid to the instability of opinion caused by influences from outside \cite{NOISE_EPJ1,NOISE_EPJ2,NOISE_EPJ3}. Hence,  the simulation in most BC models ends when static configuration is reached (consensus, polarization, or clusterization) and nothing else happens.

It is noticeable that even in  countries with stable democracy we can observe oscillations in opinion distribution, some of them are very strong. This result cannot be obtained in simulations if only the interactions between agents is taken into account. The model described in this paper tries to incorporate influence from outside.

I understand that many external factors can influence an  agent's opinion, such as dramatic event, Mass-media or others. This influence generates some unpredictable opinion changes. It may also be regarded as  free will opposing the conformist character of opinion exchange.  I assumed that the opinion of such an influenced agent may change to a completely different one, as in models with discrete opinions where it is known as "contrarians" \cite{CONT1,CONT2,CONT3,CONT4}. To simulate these various unpredictable changes noise was added to that model \cite{NOISE_EPJ1,NOISE_EPJ2,NOISE_EPJ3}. As in  \cite{NOISE_EPJ2,NOISE_EPJ3} I decided to change from time to time, an opinion of one randomly chosen agent to a new opinion, chosen from uniform random distribution between 0 and 1. Probability of this change will be described by noise parameter $\rho$ \cite{NOISE_PARAM1,NOISE_PARAM2,NOISE_PARAM3}. Another type of noise was introduced in BC models with continuous opinions \cite{DEFNOISE}, but this did not significantly affect the simulation results.

In the next section, a new BC model with continuous opinions and random changes in an agent's opinion (noise parameter $\rho$) will be introduced.

\section{Model}
\label{sec:2}

Consider a set of N agents. Each agent is connected with the others (such a structure can be described by a complete graph) and has its own opinion which is represented by a real number between 0 and 1.

\begin{enumerate}
\item Randomly choose one agent from set $A=\left\{ a_{1},...,a_{n}\right\} $ denoting it's
opinion  by $S^*$.
\item Randomly choose $L$ agents from the rest of the set. Their opinions
will be subset $\left\{ S_{i}\right\}$ of $A$ where $i=\left\{ 1,...,L\right\} $.
These agents will be listeners.
\item For each $i\epsilon\left\{ 1,...,L\right\} $ if $|S_{i}-S^*|\leq T\Rightarrow S_{i}^{'}=\frac{1}{2}(S^*+S_{i})$ .
\item With probability $\rho$, randomly choose one and only one
agent from set $A$ and change its opinion value to a new randomly chosen
value from division $\left\langle 0,1\right\rangle $.
\item Back to 1.
\end{enumerate}

As described, the parameters of the simulation are: $L$ - number of listeners,
$\rho$ - noise parameter, $T$ - tolerance and $N$ - number of agents. 

The model proposed here is quite similar to the most popular BC models,
i.e., the Deffuant (D) model \cite{DNAW00} and that of Hegselmann and Krause (HK) \cite{HK02}.
Opinions take real values in an interval $[0,1]$ and each
agent, with opinion $S^*$, interacts with agents whose opinions lie in the range $[S^*-T,S^*+T]$. 
The difference is given by the update rule: chosen agent does not interact with one of its
neighbors, like in D model, but with $L$ compatible neighbors at once, similarly to HK model.
However, on contrary to HK model, opinions of $L$ neighbors are changed, instead of changing an opinion of chosen agent. This means that one agent influences simultaneously $L$ compatible neighbors, instead of being influenced by them. Differences between these three models can be viewed shortly in the following way (see also \cite{CFL07}):
\begin{itemize}
\item        
Deffuant's model describes the opinion dynamics of large populations,
where people meet in small groups, like pairs. 
\item
HK model describes formal meetings, where there is an effective interaction involving many people at the same time.
\item
The model proposed in this paper is intended to describe how the single person can influence at the same time (during a formal meeting) a group of several listeners.
  
\end{itemize}

As mentioned in the Introduction, to describe the various unpredictable changes, the noise was introduced to the model just like Pineda et al \cite{NOISE_EPJ3} did with the Deffuant model.


It's very important that results obtained with Deffuant model  with noise are the same as with this model for $L=2$. However there's slight difference between update rules but it has no impact on the MC results, and has nothing to do with analytical approach. Hence model described in this paper could be treated as generalisation of Deffuant model with number of interacting at one time persons as a parameter.


Due to the noise, the system never reaches the final fixed point, but rather dynamic equilibrium.  Moreover, after some time the opinion distribution is independent of the initial conditions unlike in noiseless BC models case -- if there are any two different initial distributions, for example ($Q$ (e.q. uniform) and $R$ (e.q. normal)), and  $\rho_Q=\rho_R>0$, $L_Q=L_R$, $T_Q=T_R$, $N_Q=N_R$, after some number of steps we cannot distinguish between these two systems. Surely their distributions in any given moment will be different, but if a certain timespan is given, their statistical properties will be the same. All the simulations were made after the system reached dynamic equilibrium.

\section{Results}
\subsection{Monte Carlo simulations}
\label{sec:3}

\begin{figure} [!ht] 
\begin{center}
\includegraphics{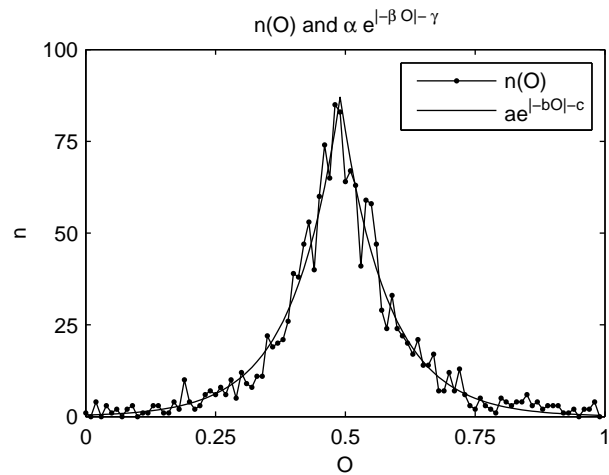}
\caption{Opinion distribution $n(O)$ could be approximated by $f\left(O\right)\approx \alpha e^{\left|-\beta O\right|- \gamma}$ where $\alpha$, $\beta$ and $\gamma$ are the factors.}
\label{fig.opinion_shape}
\end{center}
\end{figure}

\begin{figure} [!ht] 
\begin{center}
\includegraphics{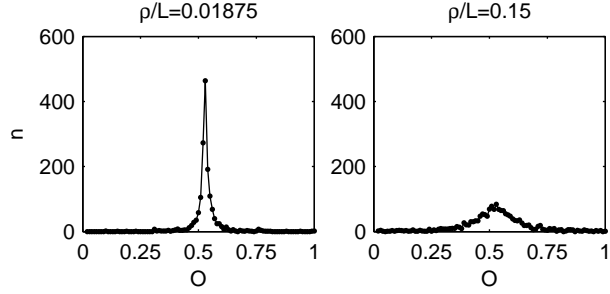}
\caption{Different shapes of opinion distribution $n(O)$ for different $\frac{\rho}{L}$. For smaller $\frac{\rho}{L}$ maximum is higher and standard deviation of distribution is greater than for greater $\frac{\rho}{L}$ ($\rho$ - noise parameter and $L$ - listeners number).}
\label{fig.different_ml}
\end{center}
\end{figure}

\begin{figure} [!ht] 
\begin{center}
\includegraphics{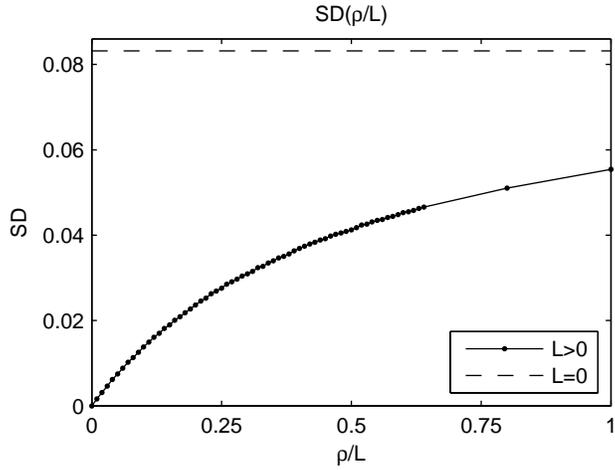}
\caption{Standard deviation of opinion distribution $SD(\frac{\rho}{L})$ for $N=1600$ and $T=1$.}
\label{fig.SD_ML}
\end{center}
\end{figure}

\begin{figure} [!ht] 
\begin{center}
\includegraphics{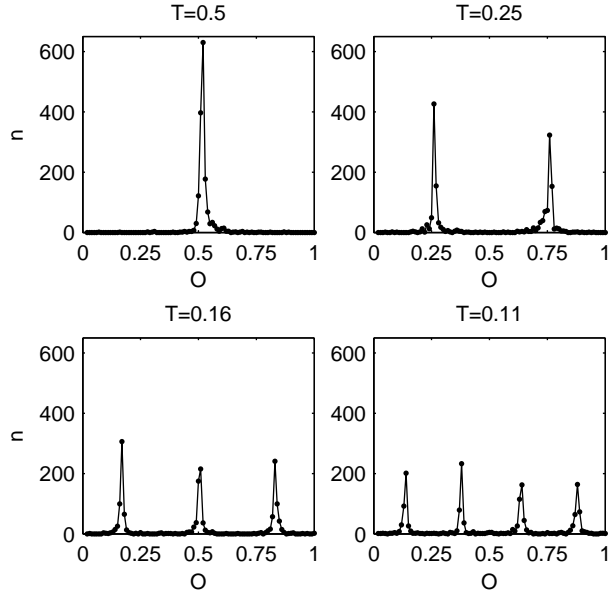}
\caption{Number of clusters in opinion space (in other words, number of modes in opinion distribution) for several values of the tolerance factor $T$, $L=16$, $\rho=0.16$, $N=1600$. }
\label{fig.clusterization}
\end{center}
\end{figure} 

\begin{figure} [!ht] 
\begin{center}
\includegraphics{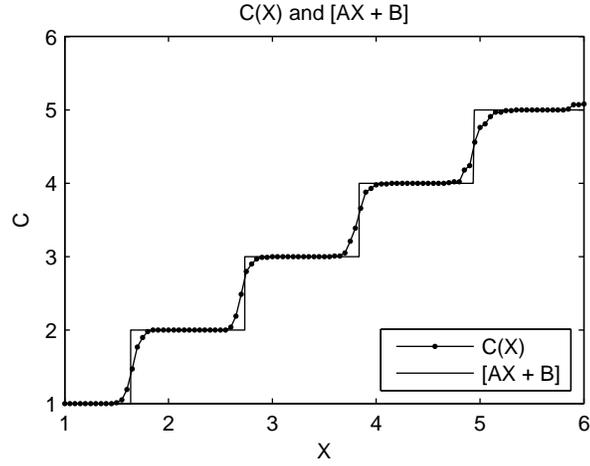}
\caption{Number of opinion clusters for $L=64$, $\rho=0.64$, $N=1600$, $10^3 MCS$, $10$ simulations, as a function
of inverse opinion threshold $X=\frac{1}{2T}$ and $C\approx\left[AX+B\right]$ approximation.}
\label{fig.C_X}
\end{center}
\end{figure}

\begin{figure}[!ht] 
\begin{center}
\includegraphics{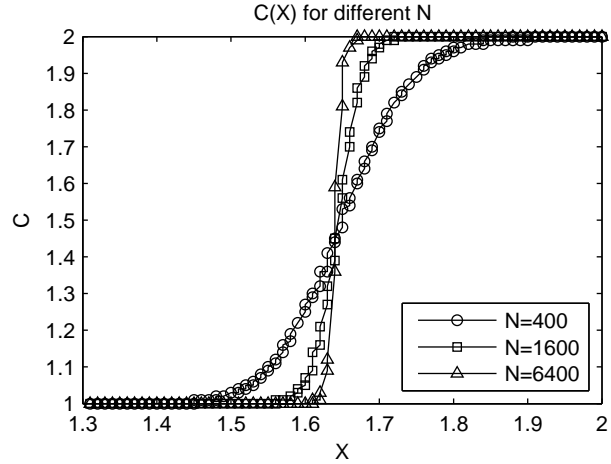}
\caption{Number of large opinion clusters  $C(X)$ for different number of agents, only first unstable region, $L=16$, $\rho=0.16$, $10^4 MCS$, $10$ simulations.}
\label{fig.C_N_first}
\end{center}
\end{figure}

\begin{figure} [!ht] 
\begin{center}
\includegraphics{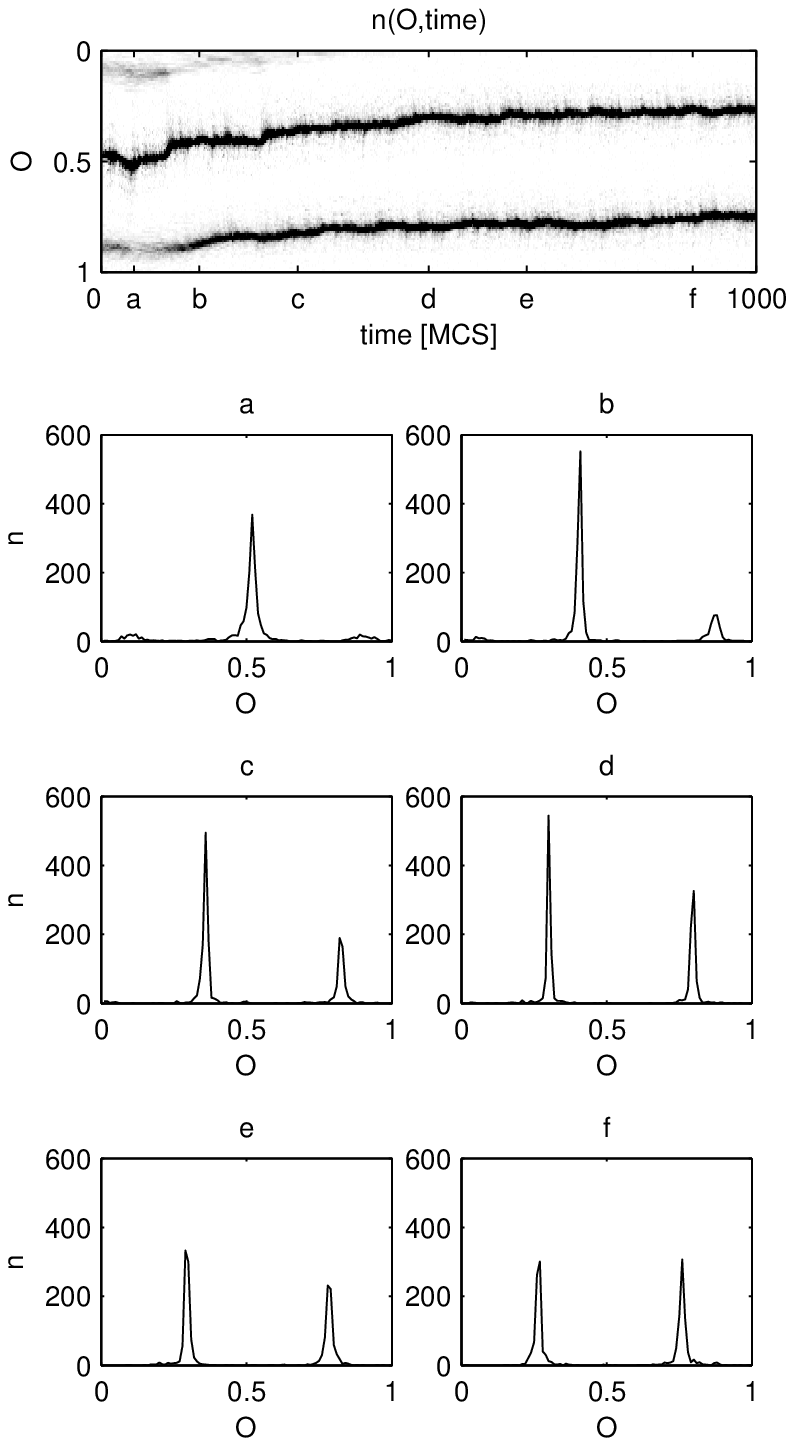}
\caption{$|1>\rightarrow|2>$  transition $L=16$, $\rho=0.16$, $X=1.66$, $N=1600$. Points a, b, c, d, e, f on upper panel of figure correspond to the six bottom panels, respectively. }
\label{fig.transition_1_2}
\end{center}
\end{figure}

\begin{figure} [!ht] 
\begin{center}
\includegraphics{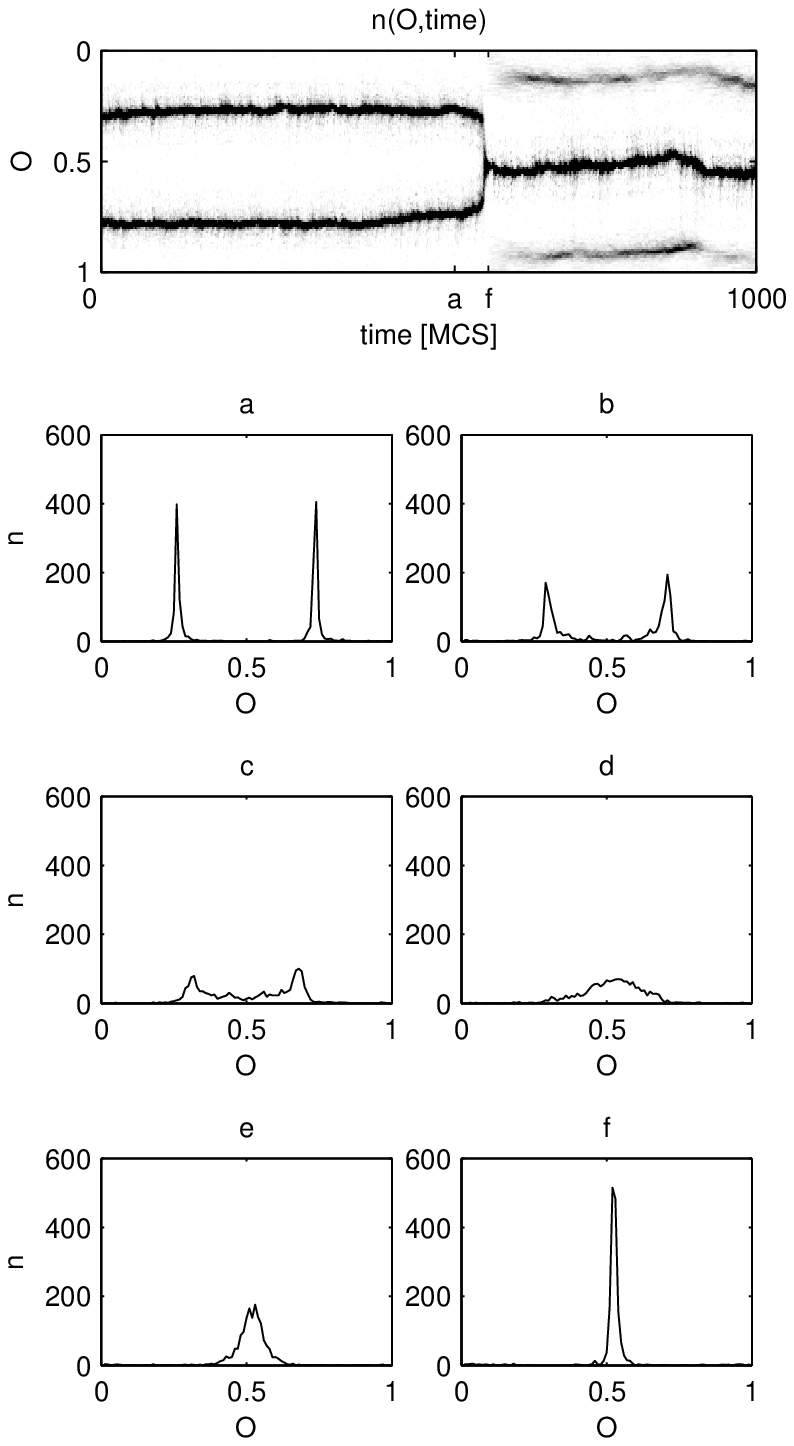}
\caption{$|2>\rightarrow|1>$  transition $L=16$, $\rho=0.16$, $X=1.66$, $N=1600$. Points a, b, c, d, e, f on upper panel of figure correspond to the six bottom panels, respectively.}
\label{fig.transition_2_1}
\end{center}
\end{figure}

\begin{figure} [!ht] 
\begin{center}
\includegraphics{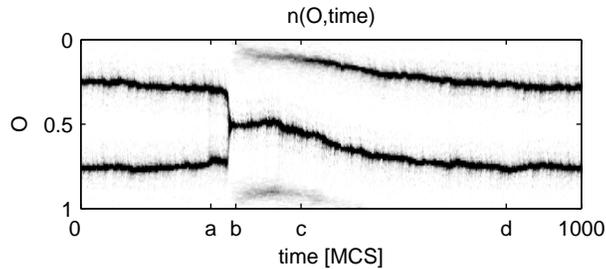}
\caption{Fragment of typical simulation outcome $|2>\rightarrow|1>$  transition in points a, b and $|1>\rightarrow|2>$  transition in points c, d $L=16$, $\rho=0.16$, $X=1.66$, $N=1600$.}
\label{fig.transition_2_1_2}
\end{center}
\end{figure}

\begin{figure} [!ht] 
\begin{center}
\includegraphics{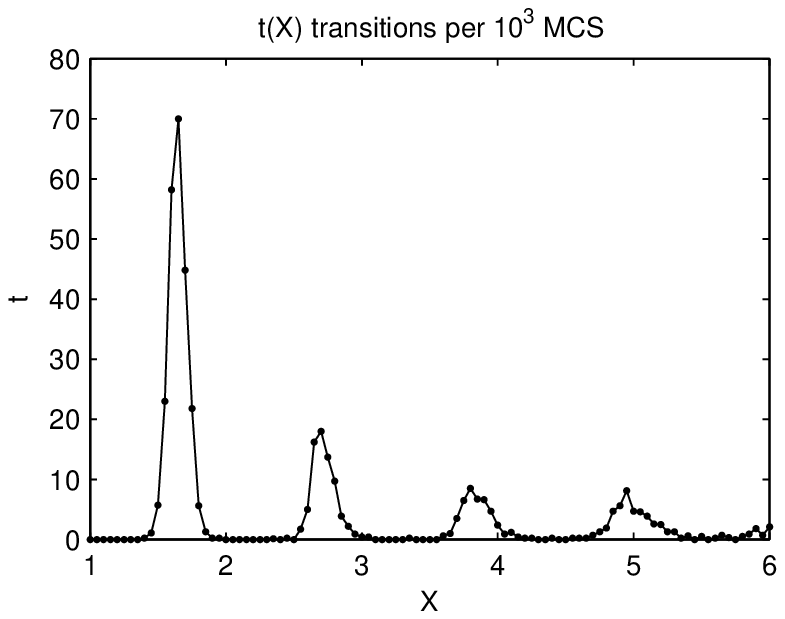}
\caption{Number of transitions $t(X)$ per $10^3 MCS$ as a function of inverse opinion threshold $X=\frac{1}{2T}$ $L=64$, $\rho=0.64$, $N=1600$, $10^3 MCS$ 10 simulations.}
\label{fig.t_X}
\end{center}
\end{figure}

\begin{itemize}
\item If $L>0$, $\rho=0$, and $T=1$, the consensus is reached very quickly.
\item If $L=0$, $\rho>0$, and $T\epsilon<0,1>$, there is no information exchange and uniform random distribution of opinions appears.
\item If $L>0$, $\rho>0$, and $T=1$, the set of agents does not reach consensus but rather stays in dynamic equilibrium with one big opinion cluster.
\end{itemize}

For $T=1$, distribution of the opinions can be approximated by:

\begin{equation}
\label{eq.opinion_shape}
f\left(O\right)\approx \alpha e^{\left|-\beta O\right|- \gamma}
\end{equation}

rule, where $\alpha$, $\beta$ and $\gamma$ are the factors that depend on model parameters (see fig.~\ref{fig.opinion_shape}).
For $T=1$ and $N=const.$, standard deviation ($SD$) of opinions depends only on $\frac{\rho}{L}$ -- the greater is factor $\frac{\rho}{L}$ the greater is $SD$ (see figs.~\ref{fig.different_ml} and fig.~\ref{fig.SD_ML}).
As it  can be seen, regardless of the force attracting the agents to the center, there are still some agents spread on the whole opinion space.

For $T<1$, as in other BC models, opinion fragmentation occurs. The smaller $T$ is, the greater number of clusters occur (see fig.~\ref{fig.clusterization}). Assuming $X=\frac{1}{2T}$, the number of large opinion clusters $C$ corresponds to
\begin{equation}
\label{eq.C_X}
C\approx\left[AX+B\right]
\end{equation}

rule (see fig.~\ref{fig.C_X}), which is more accurate than 

\begin{equation}
\label{eq.C_X_Deffuant}
C\approx\left[X\right]
\end{equation}

proposed by Deffuant \cite{DNAW00}.  
Using $X$ rather than $T$ where $X=\frac{1}{2T}$  is more suitable for the presentation of simulation results, so it is more often used in this paper. Most importantly the number of clusters $C$ changes continuously, not discretely but there are evident steps. There are regions where $C$ is far from the integer. These regions in the space of $X$ are most interesting because criticality appears there, according to bifurcation points in \cite{NOISE_EPJ3}. Due to instability in these regions,  spontaneous transitions between states with different numbers of big opinion clusters appear.

\begin{figure}[!ht] 
\begin{center}
\includegraphics{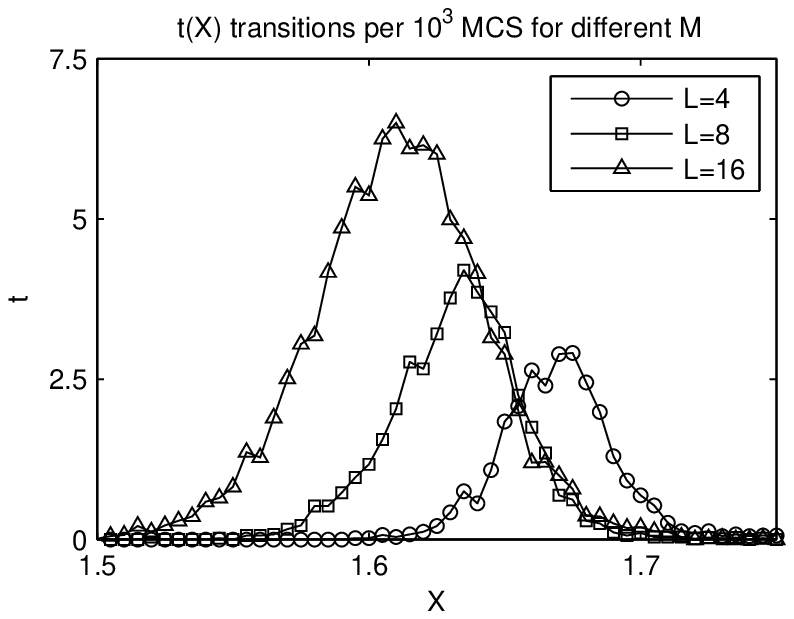}
\caption{Number of transitions per $10^3 MCS$, $t(X)$ for different number of listeners $L$, only first unstable region, $\rho=0.08$, $N=1600$, $10^4 MCS$ 10 simulations.}
\label{fig.t_L_first}
\end{center}
\end{figure}

\begin{figure}[!ht] 
\begin{center}
\includegraphics{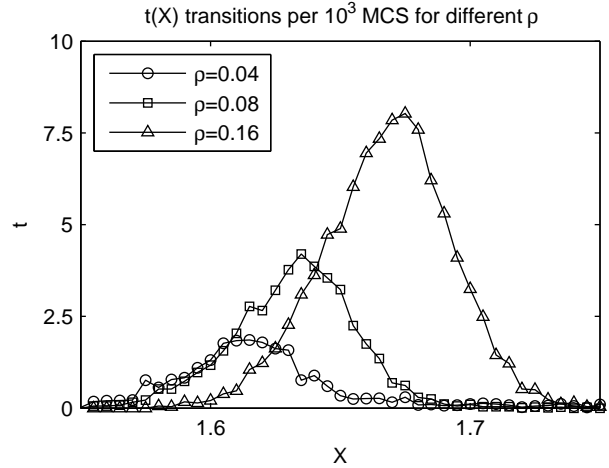}
\caption{Number of transitions per $10^3 MCS$, $t(X)$ for different noise parameter $\rho$, only first unstable region, $L=8$, $N=1600$, $10^4 MCS$ 10 simulations.}
\label{fig.t_M_first}
\end{center}
\end{figure}

\begin{figure}[!ht] 
\begin{center}
\includegraphics{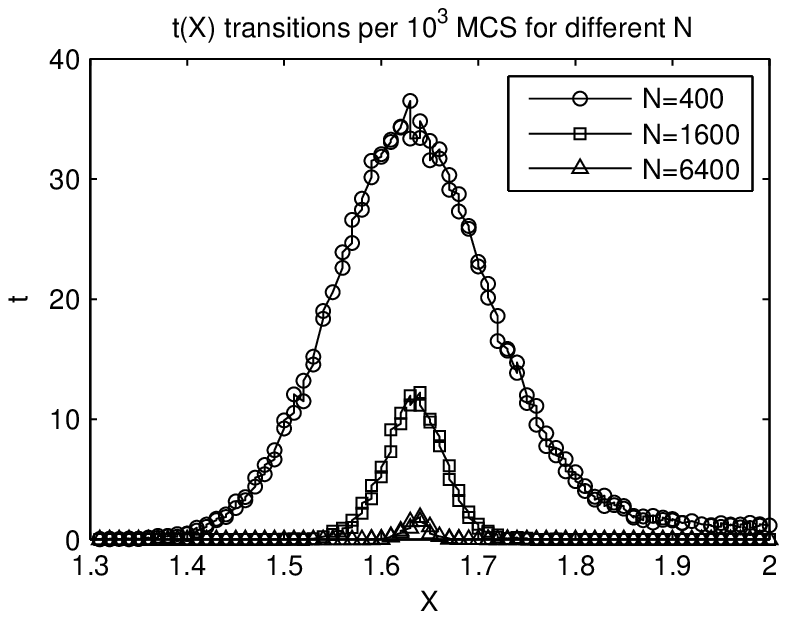}
\caption{Number of transitions per $10^3 MCS$, $t(X)$ for different number of agents, only first unstable region, $L=16$, $\rho=0.16$, $10^4 MCS$ 10 simulations.}
\label{fig.t_N_first}
\end{center}
\end{figure}

\begin{figure}[!t] 
\begin{center}
\includegraphics{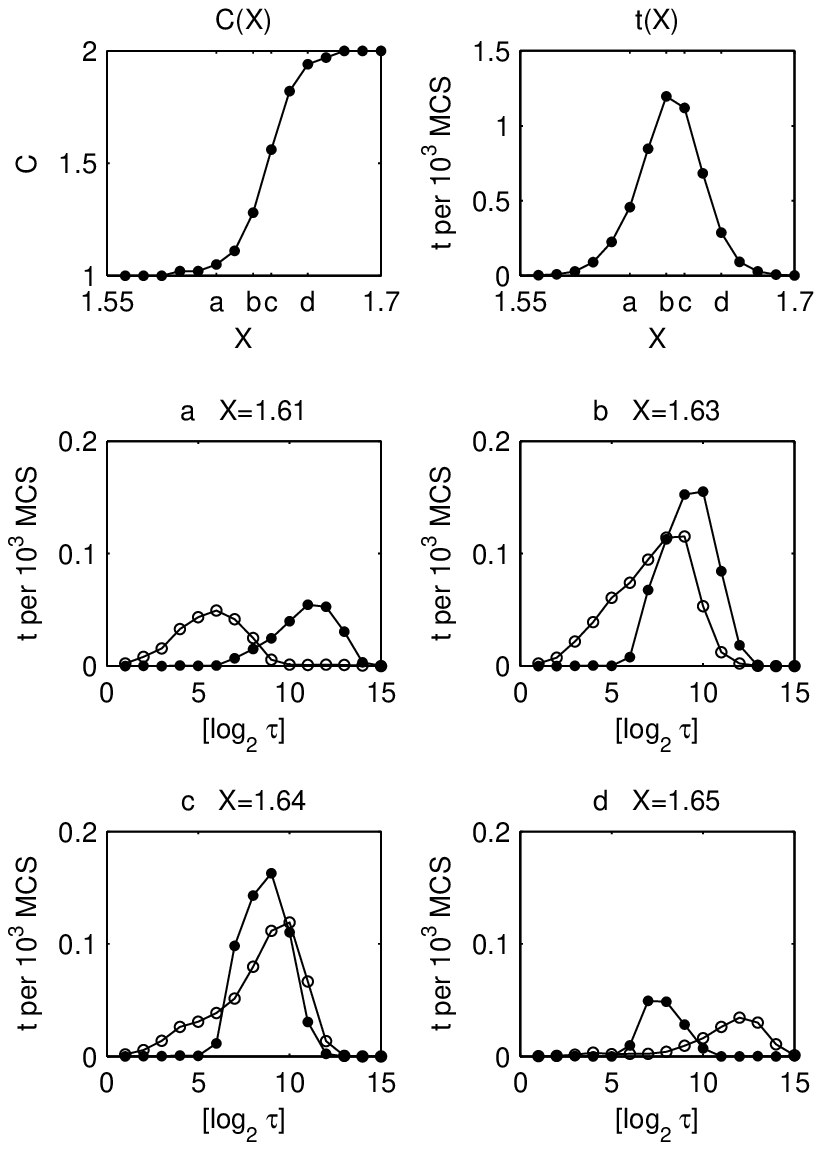}
\caption{Distribution of transitions  for parameters $L=4$, $\rho=0.04$, $N=1600$. Results for only one simulation are presented and in this case observation time is $10^7 MCS$. Disc line is $\tau_{|1>}$ distribution where $\tau_{|1>}$  is $|1>$ state lifetime or time between $|2>\rightarrow|1>$ and $|1>\rightarrow|2>$  transitions, circle line is $\tau_{|2>}$ distribution where $\tau_{|2>}$  is $|2>$ state lifetime or time between $|1>\rightarrow|2>$ and $|2>\rightarrow|1>$  transitions.}
\label{fig.transitions_distribution}
\end{center}
\end{figure}

The Deffuant model for continuous opinion dynamics under the presence of noise has been studied recently \cite{NOISE_EPJ3}. They were able to derive a master equation for the probability density function which determines the individuals density or distribution in the opinion space. Moreover, they have also found that in the noisy case the asymptotic steady-state probability distributions reached by Monte-Carlo simulations might not coincide with the ones obtained from the master equation \cite{NOISE_EPJ3}. This takes place for finite systems and is caused by perturbation introduced by noise.

Observed deviations were more pronounced in the case of being close to a bifurcation point. 

In this paper we study a new model, which differs slightly from the Deffuant model, yet belongs to the class of bounded continence models. It occurs that different BC models without noise exhibit very similar behavior \cite{CFL07,L07}. Therefore, one could expect similar behavior under the presence of noise. Indeed results obtained using Monte Carlo simulations for D model with noise \cite{NOISE_EPJ3} agrees with results obtained in this paper. Moreover, due to the similarities between BC models, results obtained here suggest that spontaneous transitions appearing in a bifurcation point might be responsible for the inconsistency between analytical results and simulations observed in \cite{CFL07,L07}.  

Let us now describe steady states of our system more carefully. For the description of the system's state, $|k>$ notation will be used, where $k$ denotes the number of big opinion clusters.

\begin{itemize}
\item For $X\epsilon <1,X_1 -\Delta>$, the system has one big opinion cluster which is in $|1>$ state, where $\Delta=\Delta(N,L,\rho)$ is a monotonically decreasing function of the total number of agents $N$. For $N \rightarrow \infty$ $\Delta \rightarrow 0$, which is usual behavior in the critical point (see fig.~\ref{fig.C_N_first}). 

\item For $X\epsilon <X_1 -\Delta,X_1 +\Delta>$ (first unstable region, see fig.~\ref{fig.C_N_first}), spontaneous transitions between one and two big clusters occur (See fig.~\ref{fig.transition_1_2}, fig.~\ref{fig.transition_2_1}), fig.~\ref{fig.transition_2_1_2}. It can be denoted as  $|1>\rightarrow|2>$ for one to two clusters transitions and  $|2>\rightarrow|1>$ for opposite. Finally there is $|1>\rightarrow|2>\rightarrow|1>$ cycle, where time intervals between transitions are unpredictable.

\item  For $X\epsilon <X_1 +\Delta,X_2 -\Delta>$ the system has two big clusters, where the second critical point $X = X_2$.

\item For $X\epsilon <X_2 -\Delta,X_2 +\Delta>$ (second unstable region), spontaneous transitions between two and three big clusters are observed. $|2>\rightarrow|3>\rightarrow|2>$ cycle occur. 

\item Generally, for $X\epsilon <X_{k-1} +\Delta,X_{k} - \Delta>$ opinions are fragmented into $k$ clusters ($k$-modal opinion distribution).

\item For $X\epsilon <X_k -\Delta,X_k +\Delta>$, spontaneous transitions between $k$ and $k+1$ big clusters are observed and $|k>\rightarrow|k+1>\rightarrow|k>$ cycle appears.  
\end{itemize}

It is surprising that such a simple model can simulate such a complex behavior. Once again it should be mentioned that spontaneous transitions occur only in critical regions around the bifurcation points $X\epsilon <X_k -\Delta,X_k +\Delta>$ and that for $N \rightarrow \infty$ $\Delta \rightarrow 0$, which is usual in the critical point. However, it should be noticed that in social systems $N$ can rarely be treated as infinite and thus $\Delta>0$. Within the proposed model, 'real life' takes place in the critical region. 

Let me now examine the spontaneous transition's mechanism.

\begin{itemize}
\item For $|1>$ state in the first unstable region there is one big cluster in the center and two small clusters near 0 and 1 (see panel denoted by (a) in fig. ~\ref{fig.transition_1_2}). The position of the central cluster is about 0.5 but it oscillates very strongly, and the oscillations are larger for smaller $N$ and are vanishing for  $N \rightarrow \infty$. Sometimes, when it goes far to one of the sides, the opposite small cluster grows very fast and becomes a second big cluster immediately and $|1>\rightarrow|2>$ transition takes place (see fig.~\ref{fig.transition_1_2} and fig.~\ref{fig.transition_2_1_2}).

\item $|2>\rightarrow|1>$ transition (See fig.~\ref{fig.transition_2_1}, fig.~\ref{fig.transition_2_1_2}) is different and more rapid. If the set is in $|2>$ state and it is in the first unstable region, there are two large clusters whose positions oscillate slightly. Sometimes they get so close to each other that their tails begin to interact and attract each other. As a result they become closer and closer, and finally joining into one big cluster. 

\item In the next unstable regions the mechanism is similar. In $|k>\rightarrow|k+1>$ transition, a new cluster is created between two other clusters. In $|k+>\rightarrow|k>$ transition, two adjacent clusters join into one.
\end{itemize}

Although the exact moment of spontaneous transition cannot be predicted, its frequency $t$ (average number of transitions per $10^3 MCS$)  can be measured. There are several maximums of transition frequency, exactly in the centres of unstable regions (float part of $C(X)\approx 0.5$) (See fig.~\ref{fig.t_X}). This is logical because the instability of an opinion's distribution is greatest in such places, so even a small perturbation can cause the spontaneous reorganization of opinions in the whole set. In each unstable region the shape of $t(X)$ can be alternately approximated by Gaussian distribution, and also follows the rule:

\begin{equation}
\label{eq.t_X_dX}
t(X) \approx \mu \frac{dC(X)}{dX},
\end{equation}

where $\mu=\mu(L,\rho,k,N)$. It can be seen (fig.~\ref{fig.t_X}) that $\mu$ is a decreasing function of $k$. This is understandable due to the fact that for a greater $X$ (greater $k$ and smaller $T$) an agent can interact with less number of listeners than in the case of a smaller $X$ (smaller $k$ and greater $T$), because they are out of confidence bound. One can say that effective $L$ is smaller and analogous to fig.~\ref{fig.t_L_first}: the smaller $L$ the fewer transitions. 

As mentioned above, the average number of transitions per time unit depends not only on $X$ (or $T$), but also on parameters $\rho$, $L$ and $N$, $t(L,\rho,X,N)$. It has been already shown that factor $\frac{\rho}{L}$ determines the shape of $n(O)$. On the other hand, for larger $\rho$ or $L$, the transitions occur much more often  (see fig.~\ref{fig.t_M_first} ~\ref{fig.t_L_first}). It is easy to understand why. For greater $\rho$, $\frac{\rho}{L}$ is also greater, hence the clusters are wider and it is easier for them to interact. When $L$ is greater, the fluctuations are also bigger. This is because when one agent whose opinion is quite rare speaks to many other agents, it can convince many agents to its rare opinion. As a result this rare opinion gets stronger and begins to attract many other agents. So the fluctuation grows and can cause the transition very easily. It's also worth to notice that for greater $\rho$ critical region shifts towards greater $X$ and for greater $L$ it shifts in opposite direction. This phenomena occurs due to changes in shape of $n(O)$ and for constant $\frac{\rho}{L}$  there is no shift. Of course it should be mentioned here once again that for $L=2$ this model behaves identically to Deffuant model with noise.

Of course, as mentioned above, spontaneous transitions occur only in critical regions $X\epsilon <X_k -\Delta,X_k +\Delta>$ and $\Delta=\Delta(N,L,\rho)$ is a monotonically decreasing function of the system size $N$ (see fig.~\ref{fig.C_N_first}). Such behavior is typical for critical phase transitions. Dependence on $N$ is also visible in $t(X)$ (see fig.~\ref{fig.t_N_first}). The more agents present, the less transitions occur. Again, this can be easily understood. With more agents, the distribution of opinion is more stable because it is hard to obtain such big fluctuations as with a small number of agents. Although for an infinite set there will not be any transitions identically as in \cite{NOISE_EPJ3} ,but social systems are finite and transitions may occur.

An analysis of the distribution of time between transitions (i.e state's lifetime) $\tau$ (see fig.~\ref{fig.transitions_distribution}) gives more interesting details. The parameters of the system were set to $L=4$, $\rho=0.04$, $N=1600$ and the observation time was $10^7 MCS$ with $1$ simulation. Simulations were provided for several different $X$ values in the first unstable region. The results are presented for $X={1.61, 1.63, 1.64, 1.65}$. It is clear that each state has its own characteristic lifetime scale -- there are maximums on the lifetime histograms. The position of the maximum depends on $X$. Where probabilities of two states are similar, the maximums of $t([\log_2(\tau)])$ distribution for higher and lower state are also close to each other (See fig.~\ref{fig.transitions_distribution} b, c). When these probabilities differ, the distributions are also further and characteristic lifetimes differs more See fig.~\ref{fig.transitions_distribution} a, d. 

\subsection{Master Equation results}

Master equation for this model when $L=1$ is:

\begin{eqnarray}
\label{eq.master_equation}
\frac{\delta P(O,t)}{\delta t} & = & (1 - \rho)[ 2\int_{|O-O'|<T/2}dO'P(2O-O',t)P(O',t) \nonumber\\
& - &  P(O,t)\int_{|O-O'|<T}dO'P(O',t) ]  +  \rho[P_{a}(O)-P(O,t)],
\end{eqnarray}

and is almost identical to that one for Deffuant system with noise derived by Pineda \cite{NOISE_EPJ3}. In fact to get that equation we need to divide first part of Pineda's equation by two. This is due to fact that in Deffuant model two agents are changing their opinions at one step but in this model for $L=1$ there's only one.

To figure out behavior of this equation I numerically integrated it using fourth order Runge-Kutta method. From analysis of stability comes out that for some values of $X$ solution of this equation is unstable. I always started from uniform opinion distribution $P(O,0)=1$ and after reaching an asymptotic solution I introduced perturbation $\varrho P_p(O)$, where $P'(O,t) = (1-\varrho)P(O,t) + \varrho P_p(O)$ was the perturbed distribution. There were two types of perturbation: symmetric $P_s(O)= 2(1 - |2O-1|)$  and asymmetric $P_a(O)= 2O$, both of them are normalized. It occurred that $|1>$ state is immune to symmetric perturbation and needs asymmetric one to make $|1>\rightarrow|2>$ transition. In $|2>$ state case situation is opposite and symmetric perturbation is needed to make $|2>\rightarrow|1>$ transition.

\begin{figure} [!ht] 
\begin{center}
\includegraphics{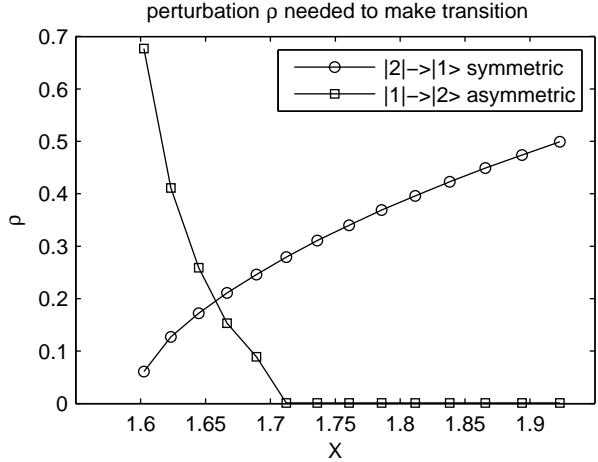}
\caption{Amount of perturbation $\rho$ needed to make transition}
\label{fig.stability}
\end{center}
\end{figure}

Therefore, in the case of $|1>$ I was introducing asymmetric perturbation $P_a$ and I was checking how big $\varrho$ has to be, to make $|1>\rightarrow|2>$ transition. If the solution was $|2>$ I was introducing symmetric perturbation $P_s$ and I was checking how big $\varrho$ has to be, to make $|2>\rightarrow|1>$ transition. The dependence between the amount of perturbation $\rho$ needed for transition and the parameter $X$ is presented in Fig.\ref{fig.stability}.  

It can be seen that for some values of $X$ stability of solution is very low and it can switch to another solution very easily but opposite switch requires much greater perturbation. However, there are values of $X$ where this switch requires perturbation of the same strength (the same $\rho$) in both directions, but still have to remember that in direction $|1>\rightarrow|2>$ we need asymmetric perturbation, and for $|2>\rightarrow|1>$ symmetric one. Because those two types of transitions are equally easy in points of equal $\rho$ their should be analogous to the maximums of $t(X)$ from MC simulations, and as can be seen in fig.\ref{fig.stability} they are very close. 


\section{Discussion}
\label{sec:4}

The model proposed in this paper is a simple bounded confidence (BC) model on a complete graph with noise $\rho$ added to simulate outside influences as well as the free will and unpredictability of individual agents. One randomly chosen agent can communicate with $L$ other randomly chosen agents according to the BC rule. Then, with probability $\rho$, the opinion of one randomly chosen agent is changed to a random number between $0$ and $1$. As in the other BC models, clusterization of opinions occurs. The new quality that appears due to the introduced noise is the presence of  spontaneous transitions between different numbers of clusters. They take place for some specific values of $X=\frac{1}{2T}$ where $T$ is tolerance.

As is usual in the case of critical phenomena, in the proposed model competition between two opposite forces is present. Due to competition between these forces, the system is in dynamic equilibrium. However, in some cases (for specific $T$ values, close to the bifurcation points) the system can be in a critical state and can spontaneously transit between two different kinds of order (i.e. states). Spontaneous transitions occur only in the critical regions $T\epsilon <T_k -\Delta,T_k +\Delta>$ and for $L \rightarrow \infty \Rightarrow \Delta \rightarrow 0$, which is usual in the critical point. However, it should be noted that in social systems $L$ can rarely be treated as infinite and thus $\Delta>0$. Within the proposed model, 'real life' takes place in the critical region. 

Occurrence of spontaneous transitions has been also observed in \cite{NOISE_EPJ3} for the Deffuant model with noise, and some analytical results were made, however no detailed analysis of this phenomena has been provided. On contrary, in this paper an analysis of the distribution of time between transitions (i.e state's lifetime) and influence of $L$ $\rho$ and $N$ on the $t(X)$, have been presented. Also influence of the perturbation on ME solutions was investigated. I hope it will shed some more light on that BC models analysis.

\section{Appendix}

Derivation of master equation for time evolution of $P(O)$ for model described in this paper is almost the same as in \cite{NOISE_EPJ3}, except one small change in number of opinions updated in one step. $P_n(O)$ is probability density function of the opinions at step n and is constructed from the histogram of individual opinions $O_n^i$. Let's choose two agents $i,j$  to update at step $n$ their opinions will be $O_n^i,O_n^j$. Probability that agent $i$ at step $n+1$ will adopt opinion $O$ is $P_{n+1}^i(O)$ and is given by below formula. 

\begin{eqnarray}
\label{eq.master_equation_step_1}
P_{n+1}^i(O) & = &  \int_{|O^i_n-O^j_n|<T/2}dO^i_ndO^j_nP_n(O^i_n)P_n(O^j_n)\delta\left(O - \frac{O^i_n + O^j_n}{2}\right)  \nonumber\\
& + &  \int_{|O^i_n-O^j_n|<T/2}dO^i_ndO^j_nP_n(O^i_n)P_n(O^j_n)\delta(O - O^i_n),
\end{eqnarray}

The independence approximation for the variables $O^i_n,O^j_n$ has been assumed and it means that $P_n(O_n^i,O_n^j)=P_n(O_n^i)P_n(O_n^j)$. To figure out $P_{n+1}(O)$ we have to include interaction between agents with probability $(1-\rho)$ as well as random change of  opinion of randomly chosen agent with probability $\rho$. For Deffuant model equation has form given below.

\begin{eqnarray}
\label{eq.master_equation_step_2}
P_{n+1}(O) & = &  (1-\rho)\left[\frac{N-2}{N}P_n(O)+\frac{1}{N}P^i_{n+1}(O)+\frac{1}{N}P^j_{n+1}(O)\right]  \nonumber\\
& + &  \rho\left[\frac{N-1}{N}P_n(O)+\frac{1}{N}P_a(O)\right] ,
\end{eqnarray}

Now we want to consider this equation for model described in this paper for case of $L=1$. Because there is only one agent changing it's opinion, the part $\frac{1}{N}P^j_{n+1}(O)$ should be neglected and $N-2$ should be replaced buy $N-1$.

\begin{eqnarray}
\label{eq.master_equation_step_3}
P_{n+1}(O) & = &  (1-\rho)\left[\frac{N-1}{N}P_n(O)+\frac{1}{N}P^i_{n+1}(O)\right]  \nonumber\\
& + &  \rho\left[\frac{N-1}{N}P_n(O)+\frac{1}{N}P_a(O)\right] ,
\end{eqnarray}

After replacing $P_{n+1}^i$ from Eq. \ref{eq.master_equation_step_1}, and simple transformations there goes:

\begin{eqnarray}
\label{eq.master_equation_step_4}
P_{n+1}(O)  =   P_n(O) + \frac{(1-\rho)}{N}[2\int_{|O-O'|<T/2}dO'P_n(2O-O')P_n(O') -  
\nonumber\\
P_n(O)\int_{|O-O'|<T}dO'P_n(O')] + \frac{\rho}{N}[P_a(O) - P_n(O)] ,
\end{eqnarray}

For continuum limit $P_n(O) \rightarrow P(O,t)$ with time $t=n\delta t$ and $\delta t = 1/N \rightarrow 0$ as $N \rightarrow \infty$ there is:

\begin{eqnarray}
\label{eq.master_equation_final}
\frac{\delta P(O,t)}{\delta t} & = & (1 - \rho)[ 2\int_{|O-O'|<T/2}dO'P(2O-O',t)P(O',t) \nonumber\\
& - &  P(O,t)\int_{|O-O'|<T}dO'P(O',t) ]  +  \rho[P_{a}(O)-P(O,t)],
\end{eqnarray}

Which is master equation for model described above for $L=1$ case, where $\rho$ denotes noise intensity.


\begin{thebibliography}{33}
\bibitem{CFL07}
C. Castellano, S. Fortunato, V. Loreto, Rev. Mod. Phys. 81, 591 (2009)
\bibitem{G08}
S. Galam, Int. J. Mod. Phys. C 19, 409 (2008)
\bibitem{SW05}
K. Sznajd-Weron, Acta Phys. Pol. B 36 2537 (2005) 
\bibitem{L07}
J. Lorenz, Int. J. Mod. Phys. C 18, 1819 (2007)
\bibitem{DNAW00}
G. Deffuant, D. Neau, F. Amblard, and G. Weisbuch, Adv. Compl. Sys. 3, 87 (2000)
\bibitem{HK02}
R. Hegselmann and U. Krause, JASSS 5(3) (2002)
\bibitem{WDA05}
G. Weisbuch, G. Deffuant and F. Amblard, Physica A 353 555 (2005)
\bibitem{CONT1}
S. Galam, Physica A 333, 453 (2004) 
\bibitem{CONT2}
M. S. de la Lama, J. M. Lopez and H. S. Wio,  Europhys. Lett. 72, 851 (2005)
\bibitem{CONT3}
C. Borghesi and S. Galam, Physical Review E 73, 066118 (2006) 
\bibitem{CONT4}
H. S. Wio, M. S. de la Lama, J. M. Lopez, Physica A 371, 108 (2006) 
\bibitem{NOISE_EPJ1}
EDMONDS, B.,(2006), Assessing the Safety of (Numerical) Representation in Social Simulation. pp. 195-214 in: Agent-based computational modelling, edited by F.C. Billari, T. Fent, A. Prskawetz and J. Schefflarn, Physica Verlag, Heidelberg 2006.
\bibitem{NOISE_EPJ2}
T. Carletti, D. Fanelli, A. Guarino, F. Bagnoli, A. Guazzini,  Eur. Phys. J. B 64, 285 (2008) 
\bibitem{NOISE_EPJ3}
M. Pineda, R. Toral and E. Hernandez-Garcia, J. Stat. Mech. P08001 (2009) 
\bibitem{NOISE_PARAM1}
K. Sznajd-Weron, J. Sznajd, Int. J. Mod. Phys. C 11, 1157 (2000)
\bibitem{NOISE_PARAM2}
F. Schweitzer and J. A. Ho{\l}yst, Eur. Phys. J. B 15, 723 (2000)
\bibitem{NOISE_PARAM3}
Medeiros, Nazareno G. F.; Silva, Ana T. C.; Moreira, F. G. Brady, Phys. Rev. E 73, 046120 (2006)
\bibitem{DEFNOISE}
G. Deffuant, JASSS  9(3) (2006)
\bibitem{SS04}
D. Stauffer, A.O. Sousa, arXiv:cond-mat/0310243v2 [cond-mat.stat-mech] (2004)
\bibitem{WDAN}
G. Weisbuch G. Deffuant , F. Amblard and J. P. Nadal, arXiv:cond-mat/0111494v1 [cond-mat.dis-nn] (2001)
\bibitem{AB02}
R. Albert and A.-L. Barab´asi, Rev. Mod. Phys. 74, 47 (2002)
\end{thebibliography}
\end{document}